\UseRawInputEncoding
\documentclass[superscriptaddress, notitlepage, reprint]{revtex4-1}
\usepackage[english]{babel}
\usepackage{amsmath,amsthm}
\usepackage{amsfonts}
\usepackage[pdfborder={0 0 0}, colorlinks=true, urlcolor=blue, linkcolor=blue, citecolor=blue]{hyperref}
\usepackage{color}
\usepackage{graphicx}
\usepackage{float}
\usepackage{subfigure}
\usepackage{lipsum}
\usepackage{epstopdf}
\usepackage{dcolumn}
\begin{document}
\title{Nonreciprocal Quantum Sensing}
\author{Dong  Xie}
\email{xiedong@mail.ustc.edu.cn}
\affiliation{College of Science, Guilin University of Aerospace Technology, Guilin, Guangxi 541004, People's Republic of China}
\author{Chunling Xu}
\affiliation{College of Science, Guilin University of Aerospace Technology, Guilin, Guangxi 541004, People's Republic of China}

\begin{abstract}
Nonreciprocity can not only generate quantum resources, but also shield noise and reverse interference from driving signals. We investigate the advantages of nonreciprocal coupling in sensing a driving signal. In general, we find that the nonreciprocal coupling
performs better than the corresponding reciprocal coupling. And we show that homodyne measurement is the optimal measurement. A single non-reciprocal coupling can increase measurement precision up to 2 times. Using $N$ non-reciprocal couplings in parallel, the measurement precision can be improved by $N^2$ times compared with the corresponding reciprocal coupling. In a non-zero temperature dissipative environment, we demonstrate that the nonreciprocal quantum sensing has better robustness to thermal noise than the reciprocal quantum sensing.
\end{abstract}
\maketitle

\textit{Introduction.-}
Quantum sensing\cite{lab1,lab2} is the use of quantum resources, such as critical phase transition points\cite{lab3,lab4,lab5}, the boundary time-crystal\cite{lab6}, coherent superposition\cite{lab7}, quantum entanglement\cite{lab8} and quantum squeezing\cite{lab9}, to break through the limits of classical measurement technology, and then develop a new generation of more precise measurement sensing technology. With the rapid development of the frontier fields, such as integrated circuits, life health, brain science, space technology, deep earth and deep sea, the quantum precision measurement technology characterized by high precision, miniaturization and low cost has gradually matured. Quantum precision measurement technology is helpful to the further development of quantum metrology\cite{lab10}. Similarly, the quantum metrology theory will further guide the quantum precision measurement technology to surpass the existing technology in terms of measurement precision, sensitivity, resolution and so on.

Quantum non-reciprocal interactions are asymmetric interactions between quantum systems, where changes in one system affect the other, but not vice versa. It can make the interaction between systems directional. The basis of nonreciprocity lies in breaking the inverse symmetry of time, which is the fundamental principle that controls the behavior of electromagnetic waves\cite{lab11,lab12,lab11a}. Nonreciprocity not only promotes one-way selectivity in the signal transmission direction, but also  shields sensitive signals from back-scattered noise. It stimulates the exploration of new functions of quantum devices\cite{lab13}. Circulators are essential components, operating as single-port couplers or isolators, which can shield the fragile quantum states of the cavity and qubits from electromagnetic noise and reflections of strong signals/pumps\cite{lab14,lab15,lab16,lab17,lab18}. In addition, there are many other applications of non-reciprocity. For example, non-reciprocal devices are used to regulate the flow of thermal noise, thus realizing thermal rectifiers in nanoscale quantum devices\cite{lab19}; non-reciprocal coupling has recently been shown to improve the energy storage efficiency of quantum batteries\cite{lab20}. As an important component of future superconducting devices, nonreciprocal superconducting electronics have been studied extensively in recent years\cite{lab21,lab22,lab23}.
The concept of quantum nonreciprocity has been widely extended to various fields, such as nonreciprocal photon blockade\cite{lab24}, which has been predicted in various systems\cite{lab25,lab26,lab27,lab28,lab29,lab30,lab31}.

Many theories and experiments have shown that non-reciprocity can produce quantum resources, such as directional entanglement\cite{lab32,lab33,lab34}, Schrodinger's cat state\cite{lab35,lab36}, quantum squeezing\cite{lab37}, nonreciprocal phase transitions\cite{lab38} and quantum correlations\cite{lab39}. These quantum resources can be used to improve measurement precision. In addition, as mentioned earlier, non-reciprocity protects quantum states from interference of noise and the driving signal in the opposite direction. Therefore, based on the above two reasons, non-reciprocity has the ability to improve the precision of parameter measurement. At present, there is a lack of systematic research on non-reciprocity in improving the quantum sensing precision.

In this letter, we hope to fill this gap and explore the advantages of non-reciprocal coupling over reciprocal coupling in quantum sensing. In terms of sensing a driving signal, we find that the nonreciprocal coupling always performs better than the reciprocal coupling. The measurement precision can be increased by up to two times through the non-reciprocal coupling. To amplify the advantage, we consider that there are $N$ non-reciprocal couplings. Compared with the reciprocal couplings, the nonreciprocal couplings can improve the measurement precision of $N^2$ scale. Finally, we find that the nonreciprocal quantum sensing can be more robust to thermal noise than the reciprocal quantum sensing.

\begin{figure}[h]
\includegraphics[scale=0.35]{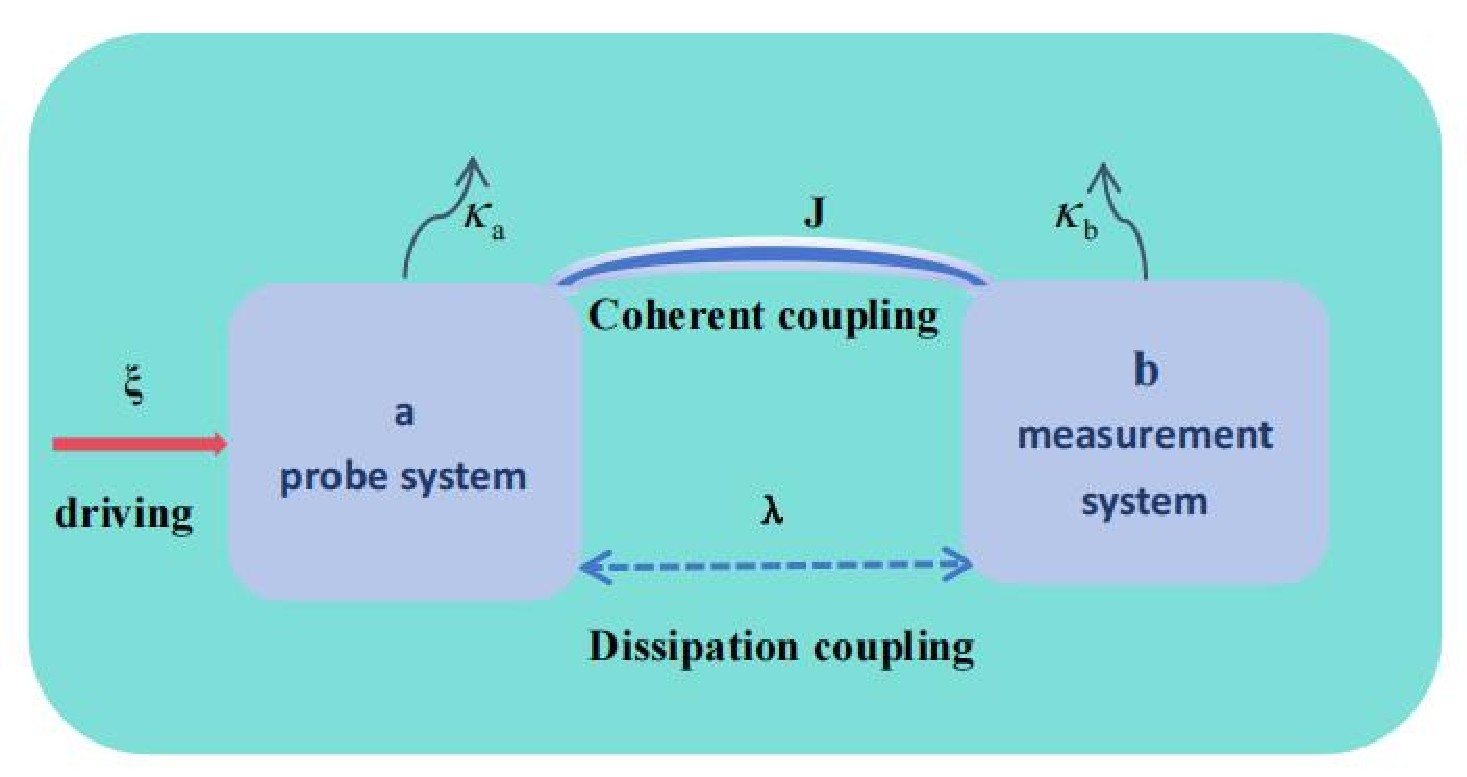}
 \caption{\label{fig.1}Schematic representation of the nonreciprocal quantum sensing system, which is composed of a probe system $a$ and a measurement system $b$.  The probe system $a$ interacts with the measurement system $b$ by a coherent coupling $J$  and a dissipation coupling $\lambda$.  The non-reciprocity is achieved by balancing the two couplings, specifically $J=i\lambda/\sqrt{2}$.
  The probe system $a$ detects a single-mode driving signal with an amplitude $\xi$, which will be read by the measurement system $b$.  $\kappa_a$ and $\kappa_b$ denote the local damping rates of each mode. }
\end{figure}

\textit{Nonreciprocal  coupling.-}We consider that a quantum sensing system is composed of a probe system with a resonance frequency of $\omega_a$ and a measurement system with a resonance frequency of $\omega_b$, as shown in Fig.~\ref{fig.1}. A driving signal with a frequency $\omega_d$ and an unknown amplitude $\xi$ directly interacts with the probe system. The driving amplitude $\xi$ denotes the parameter to be tested, which can carry the information of the magnitude of the electric filed or the power of a pump laser\cite{lab40}. The information of $\xi$  is transferred from the probe system to the measurement system by a coherent coupling with a rate $J$ and a dissipation coupling with a rate $\lambda$. Through the local measurement of the measurement system, the information of $\xi$ is finally obtained.

The Hamiltonian of the quantum sensing system is describe by($\hbar=1$)
\begin{align}
H=\omega_a a^\dagger a+\omega_b b^\dagger b+(J a^\dagger b+J^* b^\dagger a)+\xi (e^{i \omega_d t} a+e^{-i \omega_d t} a^\dagger),
\label{eq:1}
\tag{1}
\end{align}
where $a$ and $b$ are the annihilation bosonic operators of the probe system and the measurement system,  respectively.

By adiabatically eliminating the common reservoir shared by the two systems, an effective dissipative coupling can be obtained\cite{lab41,lab42}. In addition to the common bath, each system is also subject to the inevitable local dissipation. The evolution of two systems can be described by the standard master equation
\begin{align}
\dot{\rho}=-i[H,\rho]+\sum_{j=a,b}\kappa_j\mathcal{L}_j[\rho]+\lambda\mathcal{L}_z[\rho],\tag{2}
\end{align}
where the dissipation superoperator is $\mathcal{L}_o[\rho]=o\rho o^\dagger-\frac{1}{2}\{o^\dagger o, \rho\}$, $\kappa_j$ denotes the local dissipation rate, $\lambda$ denotes the nonlocal dissipation rate. We assume that the probe system and the measurement system are coupled to the common  reservoir with the same strength, i.e., the collective annihilation operator $z=(a+b)/\sqrt{2}$.

The corresponding quantum Langevin-Heisenberg equation of an operator $O$ is derived by\cite{lab43,lab44,lab45}
\begin{align}
\dot{O}=i[H,O]-\sum_{c=a,b,z}\{[O,c^\dagger](\kappa_cc-\sqrt{2\kappa_c}c_{\textmd{in}})\nonumber\\
-(\kappa_cc-\sqrt{2\kappa_c}c_{\textmd{in}})[O,c]\},
\label{eq:3}
\tag{3}
\end{align}
where $\kappa_z=\lambda$ and the expected values of noise operators $c_{\textmd{in}}=\{a_{\textmd{in}},\ b_{\textmd{in}}, z_{\textmd{in}}\}$ satisfy
\begin{align}
\langle c_{\textmd{in}}(t) \rangle=\langle c_{\textmd{in}}^\dagger(t) \rangle=0,\
\langle c^\dagger_{\textmd{in}}(t)c_{\textmd{in}} (t')\rangle=0,\label{eq:4}
\tag{4}\\
\langle c_{\textmd{in}}(t)c^\dagger_{\textmd{in}} (t')\rangle=\delta(t-t').
\label{eq:5}
\tag{5}
\end{align}
For simplicity, we assume that $\omega_a=\omega_b=\omega_d=\omega$, $\kappa=\kappa_a=\kappa_b$ and $O=\{a,\ b\}$, we obtain the detail quantum Langevin-Heisenberg equation according to Eq.~(\ref{eq:3}) in the rotating frame
\begin{align}
\dot{a}&=(-\kappa-\frac{\lambda}{\sqrt{2}})a-(\frac{\lambda}{\sqrt{2}}+iJ)b+\sqrt{2\kappa}a_{\textmd{in}}-i\xi+\sqrt{2\lambda}z_{\textmd{in}},\label{eq:6}
\tag{6}\\
\dot{b}&=(-\kappa-\frac{\lambda}{\sqrt{2}})b-(\frac{\lambda}{\sqrt{2}}+iJ^*)a+\sqrt{2\kappa}b_{\textmd{in}}+\sqrt{2\lambda}z_{\textmd{in}}.
\label{eq:7}
\tag{7}
\end{align}

When $J=i\lambda/\sqrt{2}$, the evolution of the mode $a$ is unaffected by the presence of the mode $b$.
The corresponding evolution equation of the mode $\mathbf{A}=\{a, b\}^\top$ is abbreviated to
$\mathbf{\dot{A}}=\mathbb{M}\mathbf{A}+\mathbf{A}_{\textmd{in}}$ with the evolution matrix

\[
 \mathbb{M}=\left(
\begin{array}{ll}
 -\kappa-\lambda/\sqrt{2}\ \ \ \ \ \ \ \ \ 0\\
\ \ -\sqrt{2}\lambda \ \ \ \ \ \  -\kappa-\lambda/\sqrt{2}\\
  \end{array}
\right ),\label{eq:8}
\tag{8}\]
and the noise operator $\mathbf{A}_{\textmd{in}}=\mathbf{A}^{nr}_{\textmd{in}}=(-i\xi+\sqrt{2\lambda}z_{\textmd{in}}+\sqrt{2\kappa}a_{\textmd{in}}, \sqrt{\lambda}z_{\textmd{in}}+\sqrt{2\kappa}b_{\textmd{in}})$.
The evolution matrix is completely non-reciprocal, i.e., $| \mathbb{M}_{21}|>| \mathbb{M}_{12}|=0$.

\textit{Reciprocal  coupling.-} Without the common reservoir, the master equation is described as
\begin{align}
\dot{\rho}=-i[H,\rho]+\sum_{j=a,b}\kappa_j\mathcal{L}_j[\rho].\label{eq:9}
\tag{9}
\end{align}

By the same procedure, the evolution matrix of the mode $\mathbf{A}=\{a, b\}$ is given by
\[
 \mathbb{M}^r=\left(
\begin{array}{ll}
 -\kappa\ \ \ \ \ \ \ \ \ -i J\\
-i J^*\ \ \ \ \ \  -\kappa\\
  \end{array}
\right ),\label{eq:10}
\tag{10}\]
where $J=i\lambda/\sqrt{2}=i\lambda'$.
And the noise operator $\mathbf{A}_{\textmd{in}}=(-i\xi+\sqrt{2\kappa}a_{\textmd{in}}, \sqrt{2\kappa}b_{\textmd{in}})$.
The evolution matrix $\mathbb{M}^r$ is reciprocal, i.e., $|\mathbb{M}^r_{21}|=|\mathbb{M}^r_{12}|$.

\textit{The optimal measurement.-}
For the Gaussian state, the quantum Fisher information (QFI) is derived by\cite{lab46}
\begin{align}
\mathcal{F}(\eta)=&\frac{2d^2}{4d^2+1}\textmd{Tr}[(\mathcal{C}^{-1}\partial_\xi\mathcal{C})^2]+
\frac{8(\partial_\xi d)^2}{16d^4-1}\nonumber\\
&+\langle \partial_\xi\mathbf{X}^\top\rangle \mathcal{C}^{-1} \langle \partial_\xi\mathbf{X}\rangle,\label{eq:11}
\tag{11}
\end{align}
where the abbreviation $\partial_\xi=\frac{d}{d \xi}$,
$\mathbf{X}=(q,p)^\top$ with quadrature operators defined as: $p=\frac{1}{\sqrt{2}}(b+b^\dagger)$, and $q=\frac{1}{i\sqrt{2}}(b-b^\dagger)$. And the entries of the covariance matrix are defined as $\mathcal{C}_{ij}=\frac{1}{2}\langle \mathbf{X}_i\mathbf{X}_j+\mathbf{X}_j\mathbf{X}_i\rangle-\langle {\mathbf{X}_i\rangle\langle\mathbf{X}_j}\rangle$. $d$ is given by $d=\sqrt{\textmd{Det}\mathcal{C}}$.
By the calculation (see the Supplementary Material), the QFI can be expressed as
\begin{align}
\mathcal{F}(\eta)=2|\partial_\xi \langle q\rangle|^2.\label{eq:12}
\tag{12}
\end{align}
According to the Cram\'{e}r-Rao bound\cite{lab47,lab48,lab49}, the measurement precision of $\xi$ is given by
\begin{align}
\delta \xi\geq\frac{1}{\sqrt{\mathcal{F}(\eta)}}=\frac{1}{\sqrt{2|\partial_\xi \langle q\rangle|^2}}.\label{eq:13}
\tag{13}
\end{align}

With a specific measurement operator $X$, the uncertainty of $\xi$ can be calculated by the error propagation formula
\begin{align}
\delta \xi=\frac{\sqrt{\langle X^2\rangle-\langle X\rangle^2}}{|\frac{d\langle X\rangle}{d\xi}|}.\label{eq:14}
\tag{14}
\end{align}
Using the homodyne detection with the quadrature operator $X=q$, we can get the same result as the QFI. It shows that the homodyne detection with the quadrature operator $q$ is the optimal measurement.

For the nonreciprocal quantum sensing, the estimation precision is given by
\begin{align}
\delta\xi_{nr}=\frac{(\kappa+\lambda')^2}{4\lambda'}.\label{eq:15}
\tag{15}
\end{align}

For the reciprocal quantum sensing, the estimation precision is given by
\begin{align}
\delta\xi_r=\frac{\kappa^2+\lambda'^2}{2\lambda'}.\label{eq:16}
\tag{16}
\end{align}
The ratio of measurement uncertainty is defined as
\begin{align}
\eta=\delta\xi_{nr}/\delta\xi_r=\frac{(\kappa+\lambda')^2}{2(\kappa^2+\lambda'^2)}.\label{eq:17}
\tag{17}
\end{align}
We further get
\begin{align}
1/2\leq\eta\leq1.\label{eq:18}
\tag{18}
\end{align}
It shows that the nonreciprocal quantum sensing performs better than the reciprocal quantum sensing unless $\kappa=\lambda'$. When $\kappa\gg\lambda'$ or $\kappa\ll\lambda'$, the nonreciprocity results in a two-fold increase in the measurement precision, i.e., $\eta\simeq2$.

\textit{Measurement before reaching steady state.-}
For the reciprocal Hamiltonian, the measurement uncertainty of $\xi$ at the time $t$ is
\begin{align}
\delta\xi_r=\frac{\kappa^2 + \lambda'^2}{2e^{-\kappa t} [e^{\kappa t} \lambda' - \lambda' \cos(\lambda' t) -\kappa \sin(\lambda' t)]}\label{eq:19}
\tag{19}
\end{align}
For the non-reciprocal Hamiltonian, the measurement uncertainty is
\begin{align}
\delta\xi_{nr}=\frac{(\kappa + \lambda')^2}{4 \lambda' [1 -
   e^{-t (\kappa + \lambda')} (1 + \kappa t  + \lambda't)]}\label{eq:20}
\tag{20}
\end{align}

From Eq.~(\ref{eq:20}), we can prove that the optimal precision by the nonreciprocal Hamiltonian is obtained at the steady state, i.e., $t\rightarrow\infty$.
\begin{figure}[h]
\includegraphics[scale=0.7]{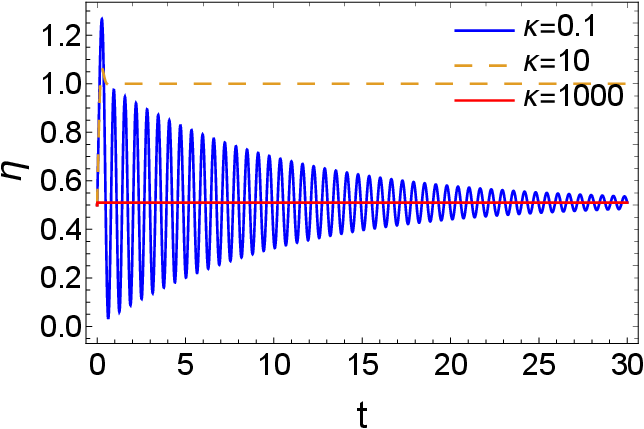}
 \caption{\label{fig.2}Evolution diagram of the ratio of measurement uncertainty $\eta$ with three decay rates. Here, the dimensionless parameters are chosen as: $\lambda=10$, $\kappa=\{0.1,1,1000\}$.}
\end{figure}
In general, the optimal precision by the reciprocal Hamiltonian is obtained before it reaches the steady state.
As shown in Fig.~\ref{fig.2}, when $\kappa\ll\lambda$ or $\kappa=\lambda$, the radio $\eta$ can be larger than 1 with unsteady state. It shows that the reciprocal Hamiltonian can perform better than the non-reciprocal Hamiltonian  in a very short period. When $\kappa\gg\lambda$, the radio $\eta$  is always less than 1. It means that the nonreciprocal quantum sensing always achieves better measurement precision than the reciprocal quantum sensing in the weak coupling.

\textit{Parallel non-reciprocal quantum sensing.- }
We consider that the mode $a$ interacts with $N$ modes $\{b_1, b_2,...,b_N\}$. The Hamiltonian in the rotating frame is described by
\begin{align}
H_N=\sum_{j=1}^N(J_p a^\dagger b_j+J_p^*a b_j^\dagger)+\xi ( a+ a^\dagger).
\label{eq:21}
\tag{21}
\end{align}
By eliminating the Markovian reservoir,  the standard master equation of the probe system $a$ and $N$  subsystems $\{b_1, b_2,...,b_N\}$ is described by
\begin{align}
\dot{\rho}=-i[H_N,\rho]+\sum_{j=a,b_1,b_2,...,b_N}\kappa_j\mathcal{L}_j[\rho]+\sum_{k=1}^N\lambda\mathcal{L}_{z_k}[\rho] .\label{eq:22}
\tag{22}
\end{align}
where the dissipation superoperator is $\mathcal{L}_o[\rho]=o\rho o^\dagger-\frac{1}{2}\{o^\dagger o ,\rho\}$, $\kappa_j$ denotes the local dissipation rate, $\lambda$ denotes the nonlocal dissipation rate. We assume that the probe system and the measurement subsystems are coupled to the common  reservoir with the same strength, i.e., the collective annihilation operator $z_k=(a+b_k)/\sqrt{2}$.
Let $\kappa=\kappa_j$ with $j=a,b_1,b_2,...,b_N$, we obtain the quantum Langevin-Heisenberg equations in the rotating frame
\begin{align}
\dot{a}=&(-\kappa-N\lambda')a-(N\lambda'+iJ_p)\sum_{j=1}^Nb_j\nonumber\\
&+\sqrt{2\kappa}a_{\textmd{in}}-i\xi+\sum_{j=1}^N\sqrt{2\lambda}z_{\textmd{in}}^j,\label{eq:23}
\tag{23}\\
\dot{b}_j=&(-\kappa-\lambda')b_j-(\lambda'+iJ_p^*)a+\sqrt{2\kappa}b_{\textmd{in}}^j+\sqrt{2\lambda}z_{\textmd{in}}^j.
\label{eq:24}
\tag{24}
\end{align}

When the coherent coupling $J_p=iN\lambda'$, the coupling between the probe system $a$ and the subsystems $b$ becomes nonreciprocal. For a long time, the system is in a steady state. The corresponding expected values are given by
\begin{align}
&\langle b_j\rangle_s=\frac{i\xi \lambda'(N+1)}{(\kappa+N\lambda')(\kappa+\lambda')},\
\langle b_j^\dagger b_j\rangle_s=\frac{\xi^2 \lambda'^2(N+1)^2}{(\kappa+N\lambda')^2(\kappa+\lambda')^2},\label{eq:25}
\tag{25}\\
&\langle b_j^2\rangle_s=\langle b_j^{2\dagger}\rangle_s=\frac{-\xi^2 \lambda'^2(N+1)^2}{(\kappa+N\lambda')^2(\kappa+\lambda')^2}.\label{eq:26}
\tag{26}
\end{align}
Utilizing the error propagation formula, the measurement uncertainty of $\xi$ is obtained by the measurement operator $\sum_{j=1}^N\frac{1}{\sqrt{2}i}(b_j-b_j^\dagger)$
\begin{align}
\delta\xi_{nr}^N=\frac{(\kappa+\lambda')(\kappa+N\lambda')}{2\sqrt{N}\lambda'(N+1)}.\label{eq:27}
\tag{27}
\end{align}
It shows that the measurements precision can be further improved by $N$ nonreciprocal couplings between measurement subsystems and the probe system.

\textit{Parallel reciprocal quantum sensing.- }Without the common reservoir, the standard master equation is described by
\begin{align}
\dot{\rho}=-i[H_N,\rho]+\sum_{j=a,b_1,b_2,...,b_N}\kappa_j\mathcal{L}_j[\rho].\label{eq:28}
\tag{28}
\end{align}
Let $\kappa=\kappa_j$ with $j=a,b_1,b_2,...,b_N$ and $J_p=iN\lambda'$, we obtain the quantum Langevin-Heisenberg equations in the rotating frame
\begin{align}
&\dot{a}=-\kappa a+\sum_{j=1}^NN\lambda'b_j+\sqrt{2\kappa}a_{\textmd{in}}-i\xi+\sum_{j=1}^N\sqrt{2\lambda}z_{\textmd{in}}^j,\label{eq:29}
\tag{29}\\
&\dot{b}_j=-\kappa b_j-N\lambda'a+\sqrt{2\kappa}b_{\textmd{in}}^j+\sqrt{2\lambda}z_{\textmd{in}}^j.\label{eq:30}
\tag{30}
\end{align}

The corresponding expected values at the steady state are given by
\begin{align}
&\langle b_j\rangle_s=\frac{i\xi N \lambda'}{\kappa^2+N^3\lambda'^2},\ \langle b_j^\dagger b_j\rangle_s=\frac{\xi^2 \lambda'^2N^2}{(\kappa^2+N^3\lambda'^2)^2},\label{eq:31}
\tag{31}\\
&\langle b_j^2\rangle_s=\langle b_j^{2\dagger}\rangle_s=\frac{-\xi^2 \lambda'^2N^2}{(\kappa^2+N^3\lambda'^2)^2}.\label{eq:32}
\tag{32}
\end{align}
Utilizing the error propagation formula, the measurement uncertainty of $\xi$ is obtained by the measurement operator $\sum_{j=1}^N\frac{1}{\sqrt{2}i}(b_j-b_j^\dagger)$
\begin{align}
\delta\xi_{r}^N=\frac{\kappa^2+N^3\lambda'^2}{2\sqrt{N}N\lambda'}.\label{eq:33}
\tag{33}
\end{align}

\textit{The advantage of parallel non-reciprocal quantum sensing.- }The radio of the parallel measurement uncertainty of the nonreciprocal quantum sensing and the reciprocal quantum sensing is given by
\begin{align}
\eta_N=\delta\xi_{nr}^N/\delta\xi_{r}^N
=\frac{N(\kappa+\lambda')(\kappa+N\lambda')}{(N+1)(\kappa^2+N^3\lambda'^2)}.\label{eq:34}
\tag{34}
\end{align}

When $N\gg1$ and $\lambda'\gg\kappa$, we obtain the radio
 \begin{align}
\eta_N
\approx\frac{\kappa+\lambda'}{N(N+1)\lambda'}\simeq\frac{1}{N^2}.\label{eq:35}
\tag{35}
\end{align}
Therefore, for sufficiently large $N$, the nonreciprocity can obtain the measurement precision of $N$ squared times higher than that obtained by the reciprocal coupling.
When $\lambda'=\kappa$, we obtain the radio
\begin{align}
\eta_N=\frac{2N}{(N^3+1)}.\label{eq:36}
\tag{36}
\end{align}
From the above equation, we can see that the nonreciprocal quantum sensing performs better than the reciprocal quantum sensing by the parallel strategy ($N>1$).

When $N^{3/2}\lambda'\ll\kappa$, we obtain the radio
\begin{align}
\eta_N
\approx\frac{N}{(N+1)}.\label{eq:37}
\tag{37}
\end{align}
In this case, the advantage will decrease with $N$. This also means that the parallel strategy has no advantage when the dissipation rate is high and the number  $N$ of non-reciprocal couplings is not large.

\textit{Non-resonant drive.- }In the previous section we considered the drive of full resonance, and we will discuss the effect of non-resonance.
We consider that  the drive is not resonant with
each local mode, i.e., $\Delta_m=\omega_m-\omega_d$ with $m=\{a,b\}$.
When $\Delta_a>0$ and $\Delta_b>0$, without loss of generality, we set $\Delta_a=\Delta_b=\Delta$.
The detail quantum Langevin-Heisenberg equation in the rotating frame
\begin{align}
&\dot{a}=(-\kappa'-\lambda')a-(\lambda'+iJ)b+\sqrt{2\kappa}a_{\textmd{in}}-i\xi+\sqrt{2\lambda}z_{\textmd{in}},\label{eq:38}
\tag{38}\\
&\dot{b}=(-\kappa'-\lambda')b-(\lambda'+iJ^*)a+\sqrt{2\kappa}b_{\textmd{in}}+\sqrt{2\lambda}z_{\textmd{in}}.
\label{eq:39}
\tag{39}
\end{align}
where $\kappa'=\kappa+i \Delta$.

For the nonreciprocal quantum sensing, the estimation precision at the steady state is given by
\begin{align}
\delta\xi_{nr}=\frac{(\kappa+\lambda')^2+\Delta^2}{4\lambda'}.\label{eq:40}
\tag{40}
\end{align}

For the reciprocal quantum sensing, the estimation precision is given by
\begin{align}
\delta\xi_{r}=\frac{\sqrt{(\kappa^2+(\lambda'-\Delta)^2)(\kappa^2+(\lambda'+\Delta)^2)}}{2\lambda'}.\label{eq:41}
\tag{41}
\end{align}
The ratio of measurement uncertainty is achieved
\begin{align}
\eta_\Delta=\delta\xi_{nr}/\delta\xi_r=\frac{(\kappa+\lambda')^2+\Delta^2}{2\sqrt{(\kappa^2+(\lambda'-\Delta)^2)(\kappa^2+(\lambda'+\Delta)^2)}}.\label{eq:42}
\tag{42}
\end{align}

When $\kappa\gg\lambda'$, we obtain $\eta_\Delta=1/2$, which is independent of the detuning $\Delta$. It shows that non-resonance does not change the advantage of non-reciprocity in the case of weak coupling $\kappa\gg\lambda'$.
In the case of $\kappa\ll\lambda'$, $\eta_\Delta=1/2$ can also be obtained for the large detuning $\Delta\gg\lambda'$ or the small detuning $\Delta\ll\lambda'$.

When $\Delta\simeq\lambda'\geq\kappa$, we derive that $\eta_\Delta\geq1$. It shows that the nonreciprocal quantum sensing can not always perform better than the reciprocal quantum sensing.

To make a small summary, in special cases non-resonance can destroy the advantages of non-reciprocity; But in most cases, the advantages of non-reciprocity are not affected by non-resonance.

When $\Delta_a>0$ and $\Delta_b<0$, without loss of generality, we set $\Delta_a=-\Delta_b=\Delta'$.
By the same procedure, the estimation precision is given by the nonreciprocal quantum sensing
\begin{align}
\delta\xi_{nr}=\frac{(\kappa+\lambda')^2+\Delta'^2}{4\lambda'}.\label{eq:43}
\tag{43}
\end{align}
For the reciprocal quantum sensing, the estimation precision is given by
\begin{align}
\delta\xi_{r}=\frac{\kappa^2+\lambda'^2+\Delta'^2}{2\lambda'}.\label{eq:44}
\tag{44}
\end{align}
The ratio of measurement uncertainty is achieved
\begin{align}
\eta_{\Delta'}=\frac{(\kappa+\lambda')^2+\Delta'^2}{2(\kappa^2+\lambda'^2+\Delta'^2)}.\label{eq:45}
\tag{45}
\end{align}
When $|\Delta'|>0$, we can prove that $\eta_{\Delta'}$ is still less than 1 no matter what the values of $\kappa$ and $\lambda'$ are. It shows that the nonreciprocal quantum sensing can still perform better than the reciprocal quantum sensing in the case of $\omega_a>\omega_d>\omega_b$.

\textit{Robustness to thermal noise.-}
We consider that subsystems are subjected to dissipative environments with non-zero temperatures.
We also obtain the quantum Langevin-Heisenberg equation in the rotating frame
\begin{align}
&\dot{a}=(-\kappa-\lambda')a-(\lambda'+iJ)b+\sqrt{2\kappa}a_{\textmd{in}}-i\xi+\sqrt{2\lambda}z_{\textmd{in}},\label{eq:46}
\tag{46}\\
&\dot{b}=(-\kappa-\lambda')b-(\lambda'+iJ^*)a+\sqrt{2\kappa}b_{\textmd{in}}+\sqrt{2\lambda}z_{\textmd{in}}.
\label{eq:47}
\tag{47}
\end{align}
where the correlation values of the thermal noise $c_{\textmd{in}}=\{a_{\textmd{in}},b_{\textmd{in}}\}$are
\begin{align}
&\langle c_{\textmd{in}}(t) \rangle=\langle c_{\textmd{in}}^\dagger(t) \rangle=0, \label{eq:48}
\tag{48}\\
&\langle c^\dagger_{\textmd{in}}(t)c_{\textmd{in}} (t')\rangle=n(\omega, T_c)\delta(t-t'),\label{eq:49}
\tag{49}\\
&\langle c_{\textmd{in}}(t)c^\dagger_{\textmd{in}} (t')\rangle=[n(\omega, T_c)+1]\delta(t-t').\label{eq:50}
\tag{50}
\end{align}
Here, $n(\omega, T_c)=(\exp[\omega/T_c]-1)^{-1}$ with $c=\{a,b\}$.
For the nonreciprocal quantum sensing, we obtain the measurement precision
\begin{align}
\delta\xi_{nr}=\frac{(\kappa+\lambda')^2}{4\lambda'}[1+\frac{4 n(\omega,T_a)\kappa\lambda^2+2n(\omega,T_b)\kappa(\kappa+\lambda')^2}{(\kappa+\lambda)^3}].\label{eq:51}
\tag{51}
\end{align}
For the reciprocal quantum sensing, the measurement precision is given by
\begin{align}
\delta\xi_r=\frac{\kappa^2+\lambda'^2}{2\lambda'}[1+\frac{n(\omega,T_a) \lambda^2+n(\omega,T_b)(2\kappa^2+\lambda^2)}{(\kappa^2+\lambda'^2)}].\label{eq:52}
\tag{52}
\end{align}

Then, the radio of the measurement precision is derived
\begin{align}
\eta=\delta\xi_{nr}/\delta\xi_r=\mu\frac{(\kappa+\lambda')^2}{2(\kappa^2+\lambda'^2)},\label{eq:53}
\tag{53}
\end{align}
where the factor $\mu$ comes from the thermal noise.
Without loss of generality, assuming that $T_a=T_b=T$, the factor $\mu$  is given by
\begin{align}
\mu=\frac{(\kappa+\lambda)^3+4 n\kappa\lambda^2+2n\kappa(\kappa+\lambda')^2}{(1+2n)(\kappa+\lambda)^3},\label{eq:54}
\tag{54}
\end{align}where $n=n(\omega,T)$.
Then, we further achieve that
\begin{align}
\mu-1=\frac{-2n\lambda(\kappa^2+\lambda^2)}{(1+2n)(\kappa+\lambda)^3}\leq0.\label{eq:55}
\tag{55}
\end{align}
From the above equation, we can see that the factor $\mu$ is still less than 1 when the temperature is not 0 ($n\neq0$).
And the difference between $\mu$ and 1 keeps increasing as $n$ increases. This indicates that the nonreciprocal quantum sensing is more robust to thermal noise than the reciprocal quantum sensing as the temperature increases.
When $\lambda=\kappa$, $\eta=\mu=1-\frac{n}{2(1+2n)}$. This shows that in the case of $\lambda=\kappa$, when the temperature is not zero, it is not the previous conclusion, but the nonreciprocal quantum sensing does better than the reciprocal quantum sensing.
When $\lambda\gg\kappa$, $\eta=\mu/2=\frac{1}{4(1+2n)}$. When $\lambda\ll\kappa$, $\eta=\mu/2=\frac{1}{2}-\frac{n\lambda}{\kappa(1+2n)}$. It shows that the thermal noise makes the result becomes different in case of $\lambda\gg\kappa$ and $\lambda\ll\kappa$.
Obviously, the non-reciprocal quantum sensing has the best advantage over the reciprocal quantum sensing in the weak dissipation case, i.e., $\lambda\gg\kappa$. In the high temperature limit ($T\rightarrow\infty$), we can obtain $\eta\rightarrow0$. This fully shows that the nonreciprocal quantum sensing has better robustness to thermal noise.

\textit{Conclusion.-}
We have investigated the role of nonreciprocal coupling in quantum sensing. Making measurements at the steady state, we obtain quantum Fisher information and prove that the homodyne measurement is the optimal measurement. In general, the nonreciprocal quantum sensing performs better than the reciprocal quantum sensing. The nonreciprocal coupling helps to improve measurement precision by up to two times. Even in non-resonant drives or non-steady state measurements, the nonreciprocal coupling performs better than the reciprocal coupling in most cases. Using $N$ nonreciprocal couplings in parallel improves the measurement precision by a factor of $N^2$ times over the reciprocal couplings, as long as $N$ is large enough. Only when the dissipation rate is high and $N$ is not large, the parallel strategy loses the advantage. Finally, we show that nonreciprocal coupling has a stronger ability to resist thermal noise interference.

Our proposed nonreciprocal quantum sensing can be feasible in the current various experimental platforms, such as, highly-tunable cavity magnonics\cite{lab50}, superconducting quantum circuits\cite{lab51}, and optomechanical circuit\cite{lab52}. It is worth exploring whether the non-reciprocity induced by kerr nonlinear\cite{lab53} can be used to improve the sensing precision.
\section*{Acknowledgements}
This research was supported by the National Natural Science Foundation of China (Grant No. 12365001 and No. 62001134), Guangxi Natural Science Foundation ( Grant No. 2020GXNSFAA159047).

\newpage

  \begin{figure*}
 \flushleft
  \normalsize
  \subsection*{Supplementary Material for: ¡°Nonreciprocal quantum sensing¡±}
\ \ \ Here, we give a specific derivation of the measurement uncertainty in different cases. For the robustness of general local thermal noise, we obtain the ratio of measurement uncertainty dependent on two local temperatures.

\subsubsection*{The solution with the nonreciprocal Hamiltonian}
 \ \ \ The corresponding evolution equation of the mode $\mathbf{A}=\{a, b\}^\top$ can be always described by the formula $\mathbf{\dot{A}}=\mathbb{M}\mathbf{A}+\mathbf{A}_{\textmd{in}}$. The general analytical solution of $\{a, b\}$ is derived by
 \begin{align}
\mathbf{A}(t)=e^{\mathbb{M} t}\mathbf{A}(0)+\int_0^te^{\mathbb{M} (t-t')}\mathbf{A}_{\textmd{in}}(t').
\tag{S1}
\label{eq:S1}
\end{align}
 \ \ \  For the nonreciprocal coupling, the evolution matrix is given by
\[
 \mathbb{M}=\left(
\begin{array}{ll}
 -\kappa-\lambda/\sqrt{2}\ \ \ \ \ \ \ \ \ 0\\
\ \ -\sqrt{2}\lambda \ \ \ \ \ \  -\kappa-\lambda/\sqrt{2}\\
  \end{array}
\right ),\tag{S2}\]
\label{eq:s2}
and the noise operator $\mathbf{A}_{\textmd{in}}=\mathbf{A}^{nr}_{\textmd{in}}=(-i\xi+\sqrt{2\lambda}z_{\textmd{in}}+\sqrt{2\kappa}a_{\textmd{in}}, \sqrt{2\lambda}z_{\textmd{in}}+\sqrt{2\kappa}b_{\textmd{in}})$.
Substituting above equation into Eq.~(\ref{eq:S1}), we achieve the solutions
\begin{align}
a(t)&=e^{-(\kappa+\lambda') t}a_0+\int_0^te^{-(\kappa+\lambda')( t-t')}(-i\xi+\sqrt{2\kappa} a_{\textmd{in}}(t')+\sqrt{2\lambda}z_{\textmd{in}}(t')),\tag{S3}\\
b(t)&=-2t \lambda' e^{-(\kappa+\lambda') t}a_0+ e^{-(\kappa+\lambda') t}b_0+\int_0^t
e^{-(\kappa+\lambda')( t-t')}[(-2t\lambda')(-i\xi+\sqrt{2\kappa} a_{\textmd{in}}(t')\nonumber\\
&+\sqrt{2\lambda}z_{\textmd{in}}(t'))+\sqrt{2\kappa} b_{\textmd{in}}(t')+\sqrt{2\lambda}z_{\textmd{in}}(t')],
\label{eq:S4}
\tag{S4}
\end{align}
where $\lambda'=\lambda/\sqrt{2}$.
After a long time, the system arrives at the steady state.
\begin{align}
a(t\rightarrow\infty)&=\int_0^\infty e^{-(\kappa+\lambda')( t-t')}(-i\xi+\sqrt{2\kappa} a_{\textmd{in}}(t')+\sqrt{2\lambda}z_{\textmd{in}}(t')),\tag{S5}\\
b(t\rightarrow\infty)&=\int_0^\infty e^{-(\kappa+\lambda')( t-t')}[(-2t\lambda')(-i\xi+\sqrt{2\kappa} a_{\textmd{in}}(t')+\sqrt{2\lambda}z_{\textmd{in}}(t'))+\sqrt{2\kappa} b_{\textmd{in}}(t')+\sqrt{2\lambda}z_{\textmd{in}}(t')].
\label{eq:S6}
\tag{S6}
\end{align}
By utilizing the expected values of noise operators $c_{\textmd{in}}=\{a_{\textmd{in}},\ b_{\textmd{in}}, z_{\textmd{in}}\}$:
\begin{align}
\langle c_{\textmd{in}}(t) \rangle=\langle c_{\textmd{in}}^\dagger(t) \rangle=0,\
\langle c^\dagger_{\textmd{in}}(t)c_{\textmd{in}} (t')\rangle=0,\
\langle c_{\textmd{in}}(t)c^\dagger_{\textmd{in}} (t')\rangle=\delta(t-t'),
\label{eq:S7}
\tag{S7}
\end{align}
we can get the expected values of the mode $b$ over the steady state
\begin{align}
\langle b(t\rightarrow\infty)\rangle=\langle b\rangle_s=\frac{2i\xi \lambda'}{(\kappa+\lambda')^2},\
\langle b^\dagger b\rangle_s=\frac{4\xi^2 \lambda'^2}{(\kappa+\lambda')^4},\
\langle b^2\rangle_s=\langle b^{2\dagger}\rangle_s=\frac{-4\xi^2 \lambda'^2}{(\kappa+\lambda')^4}.\tag{S8}\label{eq:S8}
\end{align}
\subsubsection*{The solution with the reciprocal Hamiltonian}
\ \ \  For the reciprocal coupling, the evolution matrix of the mode $\mathbf{A}=\{a, b\}^\top$ is given by
\[
 \mathbb{M}^r=\left(
\begin{array}{ll}
 -\kappa\ \ \ \ \ \ \ \ \ -i J\\
-i J^*\ \ \ \ \ \  -\kappa\\
  \end{array}
\right ).\tag{S9}\]
\label{eq:S9}
and the noise operator $\mathbf{A}_{\textmd{in}}=(-i\xi+\sqrt{2\kappa}a_{\textmd{in}}, \sqrt{2\kappa}b_{\textmd{\textmd{in}}})$.
Let the coherent coupling $J=i\lambda'$, we obtain
\begin{align}
a(t)=e^{-\kappa t}[\cos(\lambda't)a_0+\sin(\lambda't)b_0]+\int_0^te^{-\kappa ( t-t')}
[\cos(\lambda't')(-i\xi+\sqrt{2\kappa} a_{\textmd{in}}(t'))+\sqrt{2\kappa}\sin(\lambda't')b_{\textmd{in}}(t')],\tag{S10}\\
b(t)=e^{-\kappa t}[\cos(\lambda't)b_0-\sin(\lambda't)a_0]+\int_0^t
e^{-\kappa ( t-t')}[-\sin(\lambda't')(-i\xi+\sqrt{2\kappa} a_{\textmd{in}}(t'))+\sqrt{2\kappa}\sin(\lambda't')b_{\textmd{in}}(t')],
\tag{S11}\label{eq:S11}
\end{align}
where $\lambda'=\lambda/\sqrt{2}$.

  \end{figure*}

  \begin{figure*}
  \flushleft
  \normalsize
  By the same procedure, we obtain the expected value at the steady state
\begin{align}
\langle b\rangle_s=\frac{i\xi \lambda'}{\kappa^2+\lambda'^2},\
\langle b^\dagger b\rangle_s=\frac{\xi^2 \lambda'^2}{(\kappa^2+\lambda'^2)^2},\
\langle b^2\rangle_s=\langle b^{2\dagger}\rangle_s=\frac{-\xi^2 \lambda'^2}{(\kappa^2+\lambda'^2)^2}.\tag{S12}\label{eq:S12}
\end{align}
\subsubsection*{Quantum Fisher information vs. the error propagation formula}
\ \  For the Gaussian state, the quantum Fisher information(QFI) is derived by
\begin{align}
\mathcal{F}(\xi)=&\frac{2d^2}{4d^2+1}\textmd{Tr}[(\mathcal{C}^{-1}\partial_\xi\mathcal{C})^2]+
\frac{8(\partial_\xi d)^2}{16d^4-1}+\langle \partial_\xi\mathbf{X}^\top\rangle \mathcal{C}^{-1} \langle \partial_\xi\mathbf{X}\rangle,\tag{S13}
\end{align}
where $\mathbf{X}^\top=(q,p)$ with quadrature operators defined as: $p=\frac{1}{\sqrt{2}}(b+b^\dagger)$, and $q=\frac{1}{i\sqrt{2}}(b-b^\dagger)$. And the entries of the covariance matrix are defined as $\mathcal{C}_{ij}=\frac{1}{2}\langle \mathbf{X}_i\mathbf{X}_j+\mathbf{X}_j\mathbf{X}_i\rangle-\langle {\mathbf{X}_i\rangle\langle\mathbf{X}_j}\rangle$. $d$ is given by $d=\sqrt{\textmd{Det}\mathcal{C}}$.

Using Eq.~(\ref{eq:S8}) and Eq.~(\ref{eq:S12}),  we find that the covariance matrix $\mathcal{C}$ is the same for the nonreciprocal coupling and the reciprocal coupling
\[
 \mathcal{C}=\left(
\begin{array}{ll}
 1/2\ \ \ \ \ \ 0\\
\ 0 \ \ \ \ \ \  1/2\\
  \end{array}
\right ).\tag{S14}\]
Due to that $\mathcal{C}$ and $d$ are independent of the amplitude $\xi$, the QFI is simplified as
\begin{align}
\mathcal{F}(\xi)=
\langle \partial_\xi\mathbf{X}^\top\rangle \mathcal{C}^{-1} \langle \partial_\xi\mathbf{X}\rangle
=2|\partial_\xi \langle q\rangle|^2,\tag{S15}\label{eq:S15}
\end{align}
where the second equation comes from $\langle p\rangle=\frac{1}{\sqrt{2}}\langle b+b^\dagger\rangle_s=0$, which is the same for both reciprocity and non-reciprocity.

\ \ The uncertainty of $\xi$ can be calculated by the error propagation formula
\begin{align}
\delta \xi=\frac{\sqrt{\langle X^2\rangle-\langle X\rangle^2}}{|\frac{d\langle X\rangle}{d\xi}|}.\tag{S16}\label{eq:S16}
\end{align}
By using Eq.~(\ref{eq:S8}) or Eq.~(\ref{eq:S12}), we can obtain
\begin{align}
{\langle q^2\rangle-\langle q\rangle^2}=1/2.\tag{S17}\label{eq:S17}
 \end{align}
 Substituting the above equation into the error propagation formula in Eq.~(\ref{eq:S16}), we achieve
 \begin{align}
\delta \xi=\frac{\sqrt{\langle X^2\rangle-\langle X\rangle^2}}{|\frac{d\langle X\rangle}{d\xi}|}=\frac{1}{\sqrt{2|\partial_\xi \langle q\rangle|^2}}=\frac{1}{\sqrt{\mathcal{F}(\xi)}}.\tag{S18}\label{eq:S18}
\end{align}
From the above equation, we can see that the homodyne detection can get the same result as the QFI.
 It shows that the homodyne detection with the quadrature operator $q$ is the optimal measurement.

\subsubsection*{Measurement before reaching steady state}
\ \ By using Eq.~(\ref{eq:S11}) and assuming that the initial states of the both systems are vacuum states, we can obtain the expected value of $q(t)$ in the case of reciprocal quantum sensing
\begin{align}
\langle q(t) \rangle=\frac{\sqrt{2}e^{-\kappa t} [e^{\kappa t} \lambda' - \lambda' \cos(\lambda' t) -\kappa \sin(\lambda' t)]}{\kappa^2 + \lambda'^2}
\tag{S19}\label{eq:S19}
 \end{align}
And the variance is $\langle q(t)^2 \rangle-\langle q(t) \rangle^2=1/2$.
Using the error propagation formula, the measurement uncertainty of $\xi$ is obtained,
\begin{align}
\delta\xi_r=\frac{\kappa^2 + \lambda'^2}{2e^{-\kappa t} [e^{\kappa t} \lambda' - \lambda' \cos(\lambda' t) -\kappa \sin(\lambda' t)]}.\tag{S20}
\end{align}
\end{figure*}

  \begin{figure*}
  \flushleft
  \normalsize
By the same way,  we can obtain the expected value and the variance of $q(t)$  in the case of nonreciprocal quantum sensing
\begin{align}
\langle q(t) \rangle=\frac{2\sqrt{2} \lambda' [1 -
   e^{-t (\kappa + \lambda')} (1 + \kappa t  + \lambda't)]}{(\kappa + \lambda')^2},\tag{S21}\label{eq:S21}\\
\langle q(t)^2 \rangle-\langle q(t) \rangle^2=1/2.\tag{S22}\label{eq:S22}
 \end{align}
The corresponding measurement precision is given by
\begin{align}
\delta\xi_{nr}=\frac{(\kappa + \lambda')^2}{4 \lambda' [1 -
   e^{-t (\kappa + \lambda')} (1 + \kappa t  + \lambda't)]}.\tag{S23}\label{eq:S23}
\end{align}

\subsubsection*{Non-resonant drive}
\ \ After a long time, the nonreciprocal system is at the steady state, which gives the expected values like the solution in Eq.~(\ref{eq:S8})
\begin{align}
\langle b\rangle_s=\frac{2i\xi \lambda'}{(\kappa'+\lambda')^2},\tag{S24}\label{eq:S24}\\
\langle b^\dagger b\rangle_s=\frac{4\xi^2 \lambda'^2}{(\kappa'+\lambda')^4},\tag{S25}\label{eq:S25}\\
\langle b^2\rangle_s=\langle b^{2\dagger}\rangle_s=\frac{-4\xi^2 \lambda'^2}{(\kappa'+\lambda')^4}.\tag{S26}\label{eq:S26}
\end{align}
When the reciprocal system is at the steady state, we obtain the expected values like the solution in Eq.~(\ref{eq:S12})
\begin{align}
\langle b\rangle_s=\frac{i\xi \lambda'}{\kappa'^2+\lambda'^2},\tag{S27}\label{eq:S27}\\
\langle b^\dagger b\rangle_s=\frac{\xi^2 \lambda'^2}{(\kappa'^2+\lambda'^2)^2},\tag{S28}\label{eq:S28}\\
\langle b^2\rangle_s=\langle b^{2\dagger}\rangle_s=\frac{-\xi^2 \lambda'^2}{(\kappa'^2+\lambda'^2)^2}.\tag{S29}\label{eq:S29}
\end{align}
\subsubsection*{Robustness to thermal noise}

\ \  At the steady state, the expected values are achieved by the nonreciprocal coupling
\begin{align}
\langle b\rangle_s=\frac{i\xi \lambda'}{\kappa^2+\lambda'^2},\tag{S30}\label{eq:S30}\\
\langle b^\dagger b\rangle_s=\frac{\xi^2 \lambda'^2}{(\kappa^2+\lambda'^2)^2}+\frac{2 n(\omega,T_a)\kappa\lambda^2+n(\omega,T_b)\kappa(\kappa+\lambda')^2}{(\kappa+\lambda)^3},\tag{S31}\label{eq:S31}\\
\langle b^2\rangle_s=\langle b^{2\dagger}\rangle_s=\frac{-\xi^2 \lambda'^2}{(\kappa^2+\lambda'^2)^2}.\tag{S32}\label{eq:S32}
\end{align}

\ \ At the steady state, the expected values are achieved by the reciprocal coupling
\begin{align}
\langle b\rangle_s=\frac{i\xi \lambda'}{\kappa'^2+\lambda'^2},\tag{S33}\label{eq:S33}\\
\langle b^\dagger b\rangle_s=\frac{\xi^2 \lambda'^2}{(\kappa'^2+\lambda'^2)^2}+\frac{n(\omega,T_a) \lambda^2+n(\omega,T_b)(2\kappa^2+\lambda^2)}{2(\kappa^2+\lambda'^2)},\tag{S34}\label{eq:S34}\\
\langle b^2\rangle_s=\langle b^{2\dagger}\rangle_s=\frac{-\xi^2 \lambda'^2}{(\kappa'^2+\lambda'^2)^2}.\tag{S35}\label{eq:S35}
\end{align}
For the nonreciprocal quantum sensing, we obtain the measurement precision by the error propagation formula
\begin{align}
\delta\xi_{nr}=\frac{(\kappa+\lambda')^2}{4\lambda'}[1+\frac{4 n(\omega,T_a)\kappa\lambda^2+2n(\omega,T_b)\kappa(\kappa+\lambda')^2}{(\kappa+\lambda)^3}].\tag{S36}\label{eq:S36}
\end{align}
 \end{figure*}

  \begin{figure*}
  \flushleft
  \normalsize
For the reciprocal quantum sensing, the measurement precision is given by
\begin{align}
\delta\xi_r=\frac{\kappa^2+\lambda'^2}{2\lambda'}[1+\frac{n(\omega,T_a) \lambda^2+n(\omega,T_b)(2\kappa^2+\lambda^2)}{(\kappa^2+\lambda'^2)}].\tag{S37}\label{eq:S37}
\end{align}

Then, the radio of the measurement precision is derived
\begin{align}
\eta=\delta\xi_{nr}/\delta\xi_r=\mu\frac{(\kappa+\lambda')^2}{2(\kappa^2+\lambda'^2)},\tag{S38}\label{eq:S38}
\end{align}
where the factor $\mu$ is given by
\begin{align}
\mu=\frac{1+\frac{4 n(\omega,T_a)\kappa\lambda^2+2n(\omega,T_b)\kappa(\kappa+\lambda')^2}{(\kappa+\lambda)^3}}{1+\frac{n(\omega,T_a) \lambda^2+n(\omega,T_b)(2\kappa^2+\lambda^2)}{2(\kappa^2+\lambda'^2)}}.\tag{S39}\label{eq:S39}
\end{align}

 \end{figure*}


\begin{thebibliography}{9}

\vspace{3mm}

\bibitem{lab1}C. L. Degen, F. Reinhard, and P. Cappellaro, Quantum Sensing, Rev. Mod. Phys. 89, 035002 (2017).
\bibitem{lab2}J. T. Reilly, J. D. Wilson, S. B. J\"{a}ger, C. Wilson, and M. J. Holland, Optimal Generators for Quantum Sensing, Phys. Rev. Lett. 131, 150802.
\bibitem{lab3}L. Garbe, M. Bina, A. Keller, M. G. A. Paris, and Simone Felicetti, Critical Quantum Metrology with a Finite-Component Quantum Phase Transition, Phys. Rev. Lett. 124, 120504 (2020).
\bibitem{lab4}C. Hotter, H. Ritsch, and K. Gietka, Combining Critical and Quantum Metrology, Phys. Rev. Lett. 132, 060801 (2024).
\bibitem{lab5}I. Fr\'{e}rot and T. Roscilde, Quantum Critical Metrology, Phys. Rev. Lett. 121, 020402 (2018).
\bibitem{lab6}A. Cabot, F. Carollo, and I. Lesanovsky, Continuous sensing and parameter estimation with the boundary time-crystal, Phys. Rev. Lett. 132, 050801 (2024).
\bibitem{lab7}K. C. McCormick, J. Keller, S. C. Burd, D. J. Wineland, A. C. Wilson, and D. Leibfried, Quantum-enhanced sensing of a mechanical oscillator,  Nature 572, 86 (2019).
\bibitem{lab8}Y. Xia, A. R. Agrawal, C. M. Pluchar, A. J. Brady, Z. Liu, Q. Zhuang, D. J. Wilson, and Z. Zhang, Entanglement-enhanced optomechanical sensing, Nature, Nature Photon. 17, 470¨C477 (2023).
\bibitem{lab9}X. N. Feng, M. Zhang, and L. F. Wei, Beating the Standard Quantum Limit Electronic Field Sensing by Simultaneously Using Quantum Entanglement and Squeezing, Phys. Rev. Lett. 132, 220801 (2024).
\bibitem{lab10}V. Giovannetti, S. Lloyd, and L. Maccone, Quantum Metrology, Phys. Rev. Lett. 96, 010401 (2006).
 \bibitem{lab11}P. Lodahl, S. Mahmoodian, S. Stobbe, A. Rauschenbeutel, P. Schneeweiss, J. Volz, H. Pichler, and P. Zoller, Chiral quantum optics, Nature 541, 473 (2017).
 \bibitem{lab12}C. Caloz, A. Al\`{u}, S. Tretyakov, D. Sounas, K. Achouri,
and Z.-L. Deck-L\'{e}ger, Electromagnetic nonreciprocity, Phys. Rev. Appl. 10, 047001 (2018).
\bibitem{lab11a}Y. Tokura and N. Nagaosa, Nonreciprocal responses from non-centrosymmetric quantum materials, Nat. Commun. 9, 3740 (2018).
\bibitem{lab13}D. Jalas, A. Petrov, M. Eich, W. Freude, S. Fan, Z. Yu, R. Baets, M. Popovi\'{c}, A. Melloni, J. D. Joannopoulos, M. Vanwolleghem, C. R. Doerr, and H. Renner, What is-and what is not-an optical isolator. Nature Photon. 7, 579-582 (2013).
\bibitem{lab14} J. Kerckhoff, K. Lalumi\`{e}re, B. J. Chapman, A. Blais,
and K. W. Lehnert, On-chip superconducting microwave
circulator from synthetic rotation, Phys. Rev. Applied 4,
034002 (2015).
\bibitem{lab15}C. Mahoney, J. I. Colless, S. J. Pauka, J. M. Hornibrook, J. D. Watson, G. C. Gardner, M. J. Manfra, A. C. Doherty, and D. J. Reilly, On-chip microwave quantum
hall circulator, Phys. Rev. X 7, 011007 (2017).
\bibitem{lab16}B. J. Chapman, E. I. Rosenthal, J. Kerckhoff, B. A.
Moores, L. R. Vale, J. A. B. Mates, G. C. Hilton, K. Lalumi\`{e}re, A. Blais, and K. W. Lehnert, Widely tunable on-chip microwave circulator for superconducting quantum circuits, Phys. Rev. X 7, 041043 (2017).
\bibitem{lab17}X.-W. Xu and Y. Li, Optical nonreciprocity and optomechanical circulator in three-mode optomechanical systems, Phys. Rev. A 91, 053854 (2015).
\bibitem{lab18}S. Barzanjeh, M. Wulf, M. Peruzzo, M. Kalaee, P. B.
Dieterle, O. Painter, and J. M. Fink, Mechanical on-chip microwave circulator, Nature Communications 8, 953 (2017).
\bibitem{lab19}S. Barzanjeh, M. Aquilina, and A. Xuereb, Manipulating
the flow of thermal noise in quantum devices, Phys. Rev.
Lett. 120, 060601 (2018).
\bibitem{lab20}B. Ahmadi, P. Mazurek, P. Horodecki, and S. Barzanjeh, Nonreciprocal Quantum Batteries,
 Phys. Rev. Lett. 132, 210402 (2024).
\bibitem{lab21}K. Misaki and N. Nagaosa, Theory of the nonreciprocal
Josephson effect, Phys. Rev. B 103, 245302 (2021).
\bibitem{lab22}J. Chiles, E. G. Arnault, C.-C. Chen, T. F. Q. Larson, L.
Zhao, K. Watanabe, T. Taniguchi, F. Amet, and G.
Finkelstein, Nonreciprocal supercurrents in a field-free
graphene Josephson triode, Nano Lett. 23, 5257 (2023).
\bibitem{lab23}P. Virtanen, and T. T. Heikkil\"{a}, Nonreciprocal Josephson Linear Response,Phys. Rev. Lett. 132, 046002(2004).
\bibitem{lab24}A. Imamo\={g}lu, H. Schmidt, G. Woods, and  M. Deutsch, Strongly interacting photons in a nonlinear cavity, Phys. Rev. Lett. 79, 1467 (1997).
\bibitem{lab25}R. Huang, A. Miranowicz, J. Q. Liao, F. Nori, and H. Jing, Nonreciprocal Photon Blockade, Phys. Rev. Lett. 121, 153601 (2018).
\bibitem{lab26}K. Wang, Q. Wu, Y.-F. Yu, and Z.-M. Zhang,  Nonreciprocal photon blockade in a two-mode cavity with a second-order nonlinearity. Phys. Rev. A 100, 053832 (2019).
\bibitem{lab27}X. Xia, X. Zhang, J. Xu, H. Li, Z. Fu, and Y. Yang, Giant nonreciprocal unconventional
photon blockade with a single atom in an asymmetric cavity. Phys. Rev. A 104, 063713 (2021).
\bibitem{lab28}W.-J. Gu, L. Wang, Z. Yi, and L.-H. Sun,  Generation
of nonreciprocal single photons in the chiral waveguide cavity-emitter system. Phys. Rev. A 106, 043722 (2022).
\bibitem{lab29}H. Xie, L.-W. He, X. Shang, G.-W. Lin, and X.-M. Lin,
Nonreciprocal photon blockade in cavity optomagnonics,
Phys. Rev. A 106, 053707 (2022).
\bibitem{lab30}L. Bo, X.-F. Liu, C. Wang, and  T.-J. Wang, Spinning microresonator-induced chiral optical transmission. Front. Phys. 18, 12305 (2023).
\bibitem{lab31}Y. M. Liu,,  J. Cheng,, H.-F. Wang, and X. X. Yi, Simultaneous nonreciprocal conventional photon blockades of two independent optical modes by a two-level system,
Phys. Rev. A 107, 063701 (2023).
\bibitem{lab32}Y.-F. Jiao, S.-D. Zhang, Y.-L. Zhang, A. Miranowicz, L.-M. Kuang, and H. Jing, Nonreciprocal optomechanical entanglement against backscattering losses. Phys. Rev. Lett. 125,
143605 (2020).
\bibitem{lab33}Y.-F. Jiao, J.-X. Liu, Y. Li, R. Yang, L.-M. Kuang, and H. Jing, Nonreciprocal enhancement of remote entanglement between nonidentical mechanical oscillators. Phys. Rev. Applied 18, 064008 (2022).
\bibitem{lab34}L. Orr, S. A. Khan, and N. Buchholz, High-purity entanglement of hot propagating modes using nonreciprocity, PRX Quantum, 4, 020344(2023).
\bibitem{lab35}B. Li, W. Qin, Y.-F. Jiao, C.-L. Zhai, X.-W. Xu, L.-M. Kuang, H. Jing, Optomechanical Schr\"{o}dinger cat states in a cavity Bose-Einstein condensate. Fundamental Research
3, 15-20 (2023).
\bibitem{lab36}Z.-H. Li, L.-L. Zheng, Y. Wu, X.-Y. L\"{u}, Nonreciprocal generation of Schr\"{o}dinger cat state induced by topology, arXiv: 2312.10444 (2023).
\bibitem{lab37}Q. Guo, K.-X. Zhou, C.-H. Bai, Y. Zhang, G. Li, and T. Zhang, Nonreciprocal mechanical squeezing in a spinning cavity optomechanical system via pump modulation. Phys. Rev. A 108, 033515 (2023).
\bibitem{lab38}M. Fruchart, R. Hanai, and P. B. Littlewood, Non-reciprocal phase transitions, Nature, 592(7854): 363-369 (2021).
\bibitem{lab39}X. D. Lu, W. X. Cao, W. Yi, H. Shen, and Y. H. Xiao,
Nonreciprocity and quantum correlations of light transport in hot atoms via reservoir engineering, Phys. Rev. Lett. 126, 223603 (2021).
\bibitem{lab40}X. N. Feng, M. Zhang, and L. F. Wei, Beating the Standard Quantum Limit Electronic Field Sensing by Simultaneously Using Quantum Entanglement and Squeezing, Phys. Rev. Lett. 132, 220801 (2024).
 \bibitem{lab41}Y.-X. Wang, C. Wang, and A.A. Clerk, Quantum Nonreciprocal Interactions via Dissipative Gauge Symmetry, PRX quantum, 4, 010306 (2023).
\bibitem{lab42}A. Metelmann and A. A. Clerk, Nonreciprocal photon
transmission and amplification via reservoir engineering,
Phys. Rev. X 5, 021025 (2015).
\bibitem{lab43}C. Gardiner, P. Zoller, Qauntum Noise: A Handbook of Markovian and Non-Markovian Quantum Stochastic Methods with Applications to Quantum
Optics, vol. 56 (Springer, Berlin, 2004)
\bibitem{lab44} M. Reitz, C. Sommer, C. Genes, Langevin approach to quantum optics with molecules. Phys. Rev. Lett. 122, 203602 (2019)
\bibitem{lab45}D. Xie, Chunling Xu, and A. M. Wang, Quantum thermometry with a dissipative quantum Rabi system, Eur. Phys. J. Plus  137:1323 (2022).
 \bibitem{lab46}O. Pinel, P. Jian, N. Treps, C. Fabre, and D. Braun, Quantum parameter estimation using general single-mode Gaussian states. Phys. Rev. A 88, 040102(R)(2013).
  \bibitem{lab47}H. Cram\'{e}r, Mathematical Methods of Statistics (Princeton University, Princeton, 1946)
 \bibitem{lab48}C. R. Rao, Linear Statistical Inference and Its Applications (Wiley, NewYork, 1973)
 \bibitem{lab49}S. L. Braunstein, and C. M. Caves, Statistical distance and the geometry of quantum states. Phys. Rev. Lett. 72, 3439 (1994)
\bibitem{lab50}M. Kim, A. Tabesh, T. Zegray, S. Barzanjeh, and C.-M. Hu, Nonreciprocity in cavity magnonics at millikelvin temperature, arxiv: 2303.04358, (2023)
\bibitem{lab51}B. J. Chapman, E. I. Rosenthal, J. Kerckhoff, B. A.
 Moores, L. R. Vale, J. A. B. Mates, G. C. Hilton, K. La
lumi\`{e}re, A. Blais, and K. W. Lehnert, Widely tunable
 on-chip microwave circulator for superconducting quan
tum circuits, Phys. Rev. X 7, 041043 (2017).
\bibitem{lab52}K. Fang, J. Luo, A. Metelmann, M. H. Matheny, F. Mar
quardt, A. A. Clerk, and O. Painter, Generalized non
reciprocity in an optomechanical circuit via synthetic
 magnetism and reservoir engineering, Nat Phys 13, 465
 (2017).
\bibitem{lab53}Qingtian Miao, and G. S. Agarwal, Kerr nonlinearity induced nonreciprocity in dissipatively coupled resonators, Phys. Rev. Research, 6, 033020 (2024).



\end{thebibliography}
\end{document}